\documentstyle[12pt,epsf]{article}
\newcommand{\beq}{\begin{equation}}
\newcommand{\eeq}{\end{equation}}

\title{ Conditional Density Matrix:
\\ Systems and Subsystems in Quantum Mechanics }

\author{  V.V. Belokurov, O.A. Khrustalev, V.A. Sadovnichy
 \\ and O.D. Timofeevskaya \\ {\em M.V.Lomonosov Moscow State University
(MSU) } \\ {\em Moscow, 119992, Russian Federation.}
\\ {\it e-mail: khrust@goa.bog.msu.ru}}

\date{ \ \ \  }
\begin{document}
\maketitle

\begin{abstract}

A new quantum mechanical notion --- Conditional Density Matrix ---
proposed by the authors \cite{Kh}, \cite{Sol}, is discussed and is
applied to describe some physical processes.  This notion is a
natural generalization of von Neumann density matrix for such
processes as divisions of quantum systems into subsystems and
reunifications of subsystems into new joint systems. Conditional
Density Matrix assigns a quantum state to a subsystem of a
composite system under condition that another part of the
composite system is in some pure state.

\end{abstract}

\section{Introduction}
\noindent A problem of a correct quantum mechanical description of
divisions of quantum systems into subsystems and reunifications of
subsystems into new joint systems attracts a great interest due to
the present development
 of quantum communication.

Although the theory of such processes finds room in the general
scheme of quantum mechanics proposed by von Neumann
 in 1927 \cite{Neu}, even now they are often described in a fictitious    manner. For example,
the authors of classical photon teleportation experiment
\cite{Tel}  write

{\it The entangled state
contains no information on the individual particles; it only
indicates that two particles will be in the opposite states. The
important property of an entangled pair that as soon as a
measurement on one particles projects it, say, onto

$|\leftrightarrow >$ the state of the other one is determined to be
$|\updownarrow >$, and vice versa. How could a measurement on one
of the particles instantaneously influence the state of the other
particle, which can be arbitrary far away?  Einstein, among many
other distinguished physicists, could simply not accept this
"spooky action at a distance". But this property of entangled
states has been demonstrated by numerous experiments.
}

\section{ The General Scheme of Quantum Mechanics   }
\noindent

It was W.Heisenberg who in 1925 formulated a kinematic postulate
of quantum mechanics \cite {Hei}. He proposed that there exists a
connection between matrices and physical variables:
$$   variable \quad {\cal F}
       \quad \Longleftrightarrow \quad
       matrix \quad (\hat F)_{mn}.
$$
In the modern language the kinematic postulate looks like:

{\it Each dynamical variable $\cal F$ of a system $\cal S$
corresponds to a linear operator  $\hat F$ in Hilbert space  $\cal
H$}
$$  dynamical \quad variable \quad
{\cal F} \quad \Longleftrightarrow \quad linear \quad operator
  \quad {\hat F}.
$$

The dynamics is given by the famous Heisenberg's equations
formulated in terms of commutators.
$$  {d{\hat F} \over dt} \quad = \quad
     {i \over \hbar}[{\hat H}, {\hat F}].
$$

To compare predictions of the theory with experimental data it was
necessary to understand how one can determine the values of
dynamical variables in the given state. W.Heisenberg gave a
partial  answer to this problem:

  {\it If   matrix that corresponds to the dynamical variable  is
diagonal, then its diagonal elements  define possible values for
the dynamical variable, i.e. its spectrum.}
$$    (\hat F)_{mn} = f_{m}{\delta}_{mn}
  \quad \Longleftrightarrow \quad
    \lbrace f_{m} \rbrace \quad is \quad spectrum \quad {\cal F}.
$$

The general solution of the problem was given by von Neumann in
1927. He proposed the following procedure for calculation of
average values of physical variables:
$$  < {\cal F} >
    \quad = \quad
        Tr({\hat F}{\hat {\rho}}).
$$

Here  operator $\hat \rho$  satisfies three conditions:
$$   1) \quad {\hat \rho}^{+} \quad = \quad
         {\hat \rho},
$$
$$   2) \quad Tr{\hat \rho} \quad = \quad 1,
$$
$$   3) \quad  \forall \psi \in {\cal H} \quad
   <\psi|{\hat \rho}\psi>  \quad \geq 0.
$$

By the formula for average values von Neumann found out the
correspondence  between linear operators $\hat \rho$ and states of
quantum systems:
$$  \quad state \quad of\quad a\quad system \quad \rho
    \quad \Longleftrightarrow  \quad
  linear \quad operator \quad {\hat \rho}.
$$

In this way, the formula for average values becomes quantum
mechanical definition of the notion "a state of a system". The
operator $\hat \rho$ is called {\bf Density Matrix}.

From the relation
$$  (<{\cal F>})^{*} \quad = \quad  Tr({\hat F}^{+}{\hat \rho})
$$
one can conclude that
 Hermitian-conjugate operators  correspond
to complex-conjugate variables and Hermitian operators correspond
to real variables.
$$  {\cal F} \leftrightarrow {\hat F}
   \quad \Longleftrightarrow \quad
    {\cal F}^{*} \leftrightarrow {\hat F}^{+},
$$
$$  {\cal F}  = {\cal F}^{*}
   \quad \Longleftrightarrow \quad
        {\hat F} = {\hat F}^{+}.
$$
The real variables are  called {\it observables.}

From the properties of density matrix and the definition of
positively definite operators:
$$
     {\hat F}^{+} = {\hat F}, \quad \quad
    \forall \psi \in {\cal H}  \quad
      <\psi|{\hat F}{\psi}> \quad \geq 0,
$$
it follows that the average value of nonnegative variable is
nonnegative. Moreover, the average value of nonnegative variable
is equal to zero if and only if this variable equals zero.  Now it
is easy to give the following definition:

{\it  variable $\cal F$ has a
 definite value in the state $\rho$ if and only if its
 dispersion in the state  $\rho$ is equal to
 zero.
}

In accordance to general definition of the dispersion of an
arbitrary variable
$$  {\cal D}(A) \quad = \quad
      <A^{2}> \quad - \quad (<A>)^{2}\,,
$$
the expression for dispersion of a quantum variable $\cal F$ in the state
$\rho$ has the form:
$$   {\cal D}_{\rho}({\cal F}) \quad = \quad
         Tr({\hat Q}^{2}{\hat \rho}),
$$
where $\hat Q$ is an operator:
$$  \hat Q \quad =
\quad {\hat F} - <{\cal F}>{\hat E}.
$$
If $\cal F$ is observable then $Q^{2}$ is a positive definite
variable. It follows that the dispersion of $\cal F$ is
nonnegative. And all this makes clear the above-given definition.

Since density matrix is a positive definite operator and its trace
equals 1, we see that its spectrum is pure discrete  and it can be
written in the form
$$  {\hat \rho}   \quad = \quad
     \sum_{n}p_{n}{\hat P}_{n},
$$
where  ${\hat P}_{n}$ is a complete set of self-conjugate
projective operators:
$$  {{\hat P}_{n}}^{+} = {\hat P}_{n}, \quad
 {\hat P}_{m}{\hat P}_{n} = {\delta}_{mn}{\hat P}_{m},
    \quad   \sum_{n}{\hat P}_{n} = {\hat E}.
$$
Numbers $\lbrace p_{n} \rbrace$ satisfy the condition
$$  p_{n}^{*} = p_{n},  \quad 0 \le p_{n},   \quad
   \sum_{n}p_{n}\,Tr{\hat P}_{n}  = 1.
$$
It follows that $\hat \rho$ acts according to the formula
$$   {\hat \rho}{\Psi} \quad = \quad
      \sum_{n} p_{n} \sum_{\alpha \in {\Delta}_{n}}
     {\phi}_{n\alpha}\langle \phi_{n\alpha}|{\Psi} \rangle.
$$
The vectors $\phi_{n\alpha}$ form an orthonormal basis in the
space $\cal H$. Sets ${\Delta}_{n} = \lbrace 1,...,k_{n} \rbrace$
are defined by  degeneration multiplicities $k_n$ of eigenvalues
 $p_{n}$.

Now the dispersion of the observable $\cal F$ in the state $\rho$
is given by the equation
$$  {\cal D}_{\rho}({\cal F}) \quad = \quad
   \sum_{n} p_{n} \sum_{\alpha \in {\Delta}_{n}}
      ||{\hat Q}{\phi}_{n\alpha}||^{2}.
$$
All terms in this sum are nonnegative. Hence, if the dispersion is equal to
zero, then
$$  if \quad p_{n} \not= 0,  \quad
    then \quad {\hat Q}{\phi}_{n\alpha} = 0.
$$
Using the definition of the operator $\hat Q$, we obtain
$$  if \quad p_{n} \not= 0,  \quad
   then \quad {\hat F}{\phi}_{n\alpha} =
     {\phi}_{n\alpha}\langle F \rangle.
$$

In other words, {\it if an observable ${\cal F}$ has a
definite value in the given state ${\rho}$, then this value is
equal to one of the eigenvalues of the operator ${\hat F}$. }

 In this case we have
$$  {\hat \rho}{\hat
F}{\phi}_{n\alpha} \quad = \quad
    {\phi}_{n\alpha}p_{n}\langle {\cal F} \rangle\,,
$$
$$  {\hat F}{\hat \rho}{\phi}_{n\alpha} \quad = \quad
    {\phi}_{n\alpha}\langle {\cal F} \rangle p_{n}\,,
$$
that proves the commutativity of operators $\hat F$ and $\hat
\rho$.

It is well known, that if $\hat A$  and  $\hat B$  are commutative
self-conjugate operators, then there exists self-conjugate
operator $\hat T$ with non-degenerate spectrum such that $\hat A$
and $\hat B$ are functions of  $\hat T$:
$$  {\hat T}{\Psi} \quad = \quad
   \sum_{n\alpha}
{\phi}_{n\alpha}t_{n\alpha} \langle{\phi}_{n\alpha}|{\Psi}\rangle,
$$
$$  t_{n\alpha}^{*} = t_{n\alpha}, \quad
    t_{n\alpha} \not= t_{n^{'}{\alpha}^{'}}, \quad if \quad
         (n,{\alpha}) \neq (n^{'},{\alpha}^{'}).
$$
$$  {\hat F}{\Psi} \quad = \quad
   \sum_{n\alpha}
{\phi}_{n\alpha}f_{1}(t_{n\alpha})
\langle{\phi}_{n\alpha}|{\Psi}\rangle,
$$
$$  {\hat \rho}{\Psi} \quad = \quad
   \sum_{n\alpha}
{\phi}_{n\alpha}f_{2}(t_{n\alpha})
\langle{\phi}_{n\alpha}|{\Psi}\rangle,
$$
Suppose that  $\hat F$ is an operator with non-degenerate
spectrum; then

{\it if the observable ${\cal F}$ with non-degenerate spectrum
has a definite value in the state ${\rho}$, then it is possible to
represent the density matrix of this state as a function of the
operator ${\hat F}$. }

The operator $\hat F$ can be written in the form
$$  {\hat F} \quad = \quad
    \sum_{n}f_{n}{\hat P}_{n},
$$
$$  {{\hat P}_{n}}^{+} = {\hat P}_{n}, \quad
 {\hat P}_{m}{\hat P}_{n} = {\delta}_{mn}{\hat P}_{m},
     \quad tr({\hat P}_{n}) = 1,
    \quad   \sum_{n}{\hat P}_{n} = {\hat E}.
$$
The numbers $\lbrace f_{n} \rbrace$ satisfy the conditions
$$  f_{n}^{*} = f_{n},   \quad
   f_{n} \neq f_{n^{'}}, \quad if \quad n \neq n^{'}.
$$
We obviously have
$$  {\hat F} \quad = \quad
    \sum_{n}f_{n}{\hat P}_{n}.
$$
From
$$  \langle F \rangle   \quad = \quad
        \sum_{n} p_{n}f_{n}
   \quad =  \quad f_{N},
$$
$$  \langle F^2 \rangle   \quad = \quad
        \sum_{n} p_{n}f_{n}^{2}
   \quad =  \quad f_{N}^{2}
$$
we get
$$   p_{n} \quad = \quad {\delta}_{nN}.
$$
In this case  density matrix is a projective operator satisfying
the condition
$$   {\hat \rho}^{2} \quad =  \quad {\hat \rho}.
$$
It acts as
$$  {\hat \rho}{\Psi} \quad = \quad
     {\Psi}_{N}\langle {\Psi}_{N}|{\Psi} \rangle,
$$
where $|{\Psi} \rangle$ is a vector in Hilbert space.

The average value of an arbitrary variable in this state is equal
to
$$  \langle {\cal A} \rangle
     \quad = \quad
   \langle {\Psi}_{N}|{\hat A}{\Psi}_{N} \rangle.
$$
It is so-called {\it PURE } state. If the state is not pure it is
known as {\it mixed.}

Suppose that every vector in $\cal H$ is a square integrable
function $\Psi (x)$, where  $x$  is a set of continuous and
discrete variables. Scalar product is defined by the formula
$$   \langle \Psi|\Phi \rangle \quad = \quad
     \int dx{\Psi}^{*}(x){\Phi}(x).
$$
For simplicity  we  assume that every operator $\hat F$ in $\cal
H$ acts as follows .
$$  ({\hat F}{\Psi})(x) \quad = \quad
   \int F(x,x^{'})dx^{'}{\Psi}(x^{'}).
$$
That is for any operator $\hat F$ there is an integral kernel
$F(x,x^{'})$  associated with this operator
$$  {\hat F} \quad \Longleftrightarrow \quad F(x,x^{'}).
$$
Certainly, we may use $\delta$-function if necessary.

Now the average value of the variable $\cal F$ in the state $\rho$
is given by equation
$$ \langle {\cal F} \rangle_{\rho}
   \quad = \quad
   \int F(x,x^{'})dx^{'}{\rho}(x^{'},x)dx.
$$
Here the kernel ${\rho}(x,x^{'})$ satisfies the conditions
$$   {\rho}^{*}(x,x^{'}) \quad = \quad {\rho}(x^{'},x),
$$
$$    \int {\rho}(x,x)dx \quad = \quad 1,
$$
$$  \forall {\Psi} \in {\cal H} \quad
   \int{\Psi}(x)dx{\rho}(x,x^{'})dx^{'}{\Psi}(x^{'})  \geq 0.
$$

\section{Composite System and Reduced Density Matrix}
\noindent Suppose the variables  $x$ are divided  into two parts:
$x = \lbrace y,z \rbrace$. Suppose also that the space $\cal H$ is
a direct product of two spaces  ${\cal H}_{1}$, ${\cal H}_{2}$:
$$  {\cal H} \quad = \quad {\cal H}_{1}\otimes{\cal H}_{2}.
$$
So, there is a basis in the space $\cal Н$ that can be written in
the form
$$    {\phi}_{an}(y,z) \quad = \quad
          f_{a}(y)v_{n}(z)\,.
$$
The kernel of operator $\hat F$ in this basis looks like
$$  {\hat F} \quad \Longleftrightarrow \quad
   F(y,z;y^{'},z^{'})\,.
$$
In quantum mechanics it means that the system $S$ is a unification
of two subsystems $S_{1}$  and $S_{2}$:
$$   S \quad = \quad
S_{1} \cup S_{2}\,.
$$
The Hilbert space  $\cal H$  corresponds to the system $S$ and the
spaces ${\cal H}_{1}$ and ${\cal H}_{2}$ correspond to the
subsystems $S_{1}$ and $S_{2}$.

Now suppose that a physical variable ${\cal F}_{1}$ depends on
variables ${y}$ only. The operator that corresponds to  ${\cal
F}_{1}$ has a kernel
$$   F_{1}(y,z;y^{'},z^{'}) \quad = \quad
       F_{1}(y,y^{'}){\delta}(z - z^{'})\,.
$$

The average value of $F_{1}$ in the state $\rho$ is equal to
$$  \langle F_{1} \rangle_{\rho}
   \quad = \quad
    \int F(y,y^{'})dy^{'}{\rho}_{1}(y^{'},y)dy\,,
$$
where the kernel ${\rho}_{1}$ is defined by the formula
$$   {\rho}_{1}(y,y^{'}) \quad = \quad
            \int {\rho}(y,z;y_{'},z)dz\,.
$$
The operator ${\hat \rho}_{1}$  satisfies all the properties of
Density Matrix in $S_1$. Indeed, we have
$$   {{\rho}_1}^{*}(y,y^{'}) \quad = \quad {\rho _1}(y^{'},y)\,,
$$
$$    \int {\rho _1}(y,y)dy \quad = \quad 1\,,
$$
$$  \forall {\Psi _1} \in {\cal H_1} \quad
   \int{\Psi _1}(y)dy{\rho _1}(y,y^{'})dy^{'}{\Psi _1}(y^{'})  \geq 0\,.
$$
The operator
$$   {\hat \rho}_{1} \quad = \quad
      Tr_{2}{\hat \rho}_{1+2},
$$
 is called  {\bf Reduced Density Matrix} . Thus,
the state of the subsystem $S_1$ is defined by reduced density
matrix.

The reduced density matrix for the subsystem $S_2$ is defined
analogously.
$$   {\hat \rho}_{2} \quad = \quad
      Tr_{1}{\hat \rho}_{1+2}.
$$

Quantum states  $\rho_{1}$ and $\rho_{2}$ of subsystems are
defined uniquely by the state $\rho_{1+2}$ of the composite
system.

Suppose  the system $S$ is in a pure state then a quantum state of
the subsystem  $S_{1}$ is defined by the kernel
$$   {\rho}_{1}(y,y^{'}) \quad = \quad
        \int{\Psi}(y,z)dz{\Psi}^{*}(y^{'},z).
$$
If the function ${\Psi}(y,z)$ is the product
$$  {\Psi}(y,z) \quad = \quad
          f(y)w(z), \quad \int|w(z)|^{2}dz = 1,
$$
then subsystem $S_{1}$ is a pure state , too
$$   {\rho}_{1}(y,y^{'}) \quad = \quad
        f(y)f^{*}(y^{'}), \quad \int|f(y)|^{2}dy = 1.
$$
As it was proved  by von Neumann, it is the only case when purity
of composite system is inherited by its subsystems.

Let us consider an example of a system in a pure state having
subsystems in  mixed states. Let the wave function of composite
system be
$$  {\Psi}(y,z) \quad = \quad
   {1 \over \sqrt{2}}(f(y)w(z) \pm f(z)w(y)),
$$
where $<f|w> = 0$ and $<f|f>=<w|w>=1$. The density matrix of the
subsystem $S_{1}$ has the kernel
$$   {\rho}_{1}(y,y^{'}) \quad = \quad
     {1 \over 2}
 (f(y)f^{*}(y^{'}) + w(y)w^{*}(y^{'})).
$$
The kernel of the operator  ${{\hat \rho}_{1}}^{2}$ has the form
$$   {{\rho}_{1}}^{2}(y,y^{'}) \quad = \quad
     {1 \over 4}
 (f(y)f^{*}(y^{'}) + w(y)w^{*}(y^{'})).
$$
Therefore, the subsystem $S_{1}$ is in the mixed state. Moreover,
its density matrix is proportional to unity operator. The previous
property resolves the perplexities connected with Einstein -
Podolsky - Rosen paradox.

\section{EPR - paradox}
\noindent
   Anyway, it was Shr\"{o}dinger who introduced a term "EPR-paradox".
The authors of EPR themselves always  considered their article as
a demonstration of inconsistency of present to them quantum
mechanics rather than a particular curiosity.

The main conclusion  of the paper \cite{EPR} "Can
Quantum-Mechanical Description of Physical Reality Be Considered
Complete?" published in 1935 (8 years later then the von Neumann
book) is the statement:

  {\it ..we proved that
(1) the quantum mechanical description of reality given
by wave functions not complete or
   (2) when the operators corresponding to two physical quantities
do not commute the two quantities cannot have simultaneous reality.
Starting then with the assumption that the wave function does give
a complete description of the physical reality, we arrived at the
conclusion that two physical quantities, with noncommuting operators, can
have simultaneous reality. Thus the negation of (1) leads to negation
of only other alternative (2). We can thus focused to conclude that the
quantum-mechanical description of physical reality given by wave
function is not complete.
}

After von Neumann's works this statement appears obvious. However,
in order to clarify this point of view   completely we must
understand what is "the physical reality" in EPR. In EPR-paper the
physical reality is defined as:

{\it If, without in any way disturbing a system, we can predict
with certainty (i.e., with probability equal to unity) the value
of physical quantity, then there exists an element of physical
reality corresponding to this physical quantity. }

Such definition of physical reality is a step back as compared to
von Neumann's definition. By EPR definition, the state is actual
only when at least one observable has an exact value. This point
of view is incomplete and leads to inconsistency.

When a subsystem is separated "the loss of observables" results
directly from the definition of density matrix for the subsystem.
"The occurrence" of observables in the chosen subsystem when the
quantities are measured in another "subsidiary" subsystem can be
naturally explained in the terms of conditional density matrix.

\section{Conditional Density Matrix}
\noindent The average value of a variable with the kernel
$$
F^c (x,x' ) = F_1 (y,y')u(z)u^* (z'), \quad \int |u(z)|^2 dz =1,
$$
is equal to
 $$  \langle F^c \rangle_{\rho}
   \quad = \quad
p \int F_1 (y,y^{'})dy^{'}{\rho}^{c}(y^{'},y)dy,
$$
where
$$
{\rho}^{c}(y,y^{'}) = {1 \over p} \int u^*(z)dz\, \rho
(y,z;y',z')\,u(z')dz'\,,
$$
$$
p \quad = \quad \int u^* (z)dz \,\rho (y,z;y,z')\,u(z')dz' dy.
$$
Since we can represent $p$ in the form
$$
p\quad = \quad  \int P(z,z')dz'\, \rho _2 (z';z)dz,
$$
$$
 P(z,z') \quad = \quad u(z)u^*(z'),
$$
we see that $p$ is an average  value  of a variable $P$ of the
subsystem $S_2$. Operator $\hat P$ is a projector ($\hat P^2 =
\hat P$). Therefore  it is possible to consider the value $p$ as a
probability.

It is easy to demonstrate that the operator $\hat{\rho}^c $
satisfies all the properties of density matrix. So the kernel
$\rho^c (y,y')$ defines some state of the subsystem $S_1$. What is
this state?

According to the decomposition of $\delta$-function
$$
\delta (z-z')=\sum_{n} \phi _n (z) {\phi _n}^* (z'),
$$
$\{ \phi _n (z) \}$ being a basis in the space ${\it H}_2$, the
reduced density matrix is represented in the form of the sum
$$
\rho _1 (y,y') = \sum p_n {\rho}_{n}^{c} (y,y').
$$
Here
$$
{\rho}_{n}^{c} (y,y') = {1 \over p_n } \int {\phi _n}^* (z)dz \,
\rho (y,z;y',z')\, {\phi}_n  (z')dz'
$$
and
$$
p_n \quad = \quad \int {\phi _n}^* (z)dz \,\rho (y,z;y,z')\,{\phi
_n}(z')dz'dy
$$
$$
\quad = \quad \int \hat P_n (z,z')dz'\, {\rho}_2 (z',z)dz.
$$
The numbers $p_n$ satisfy  the conditions
$$
 {p_n}^* =p_n,\qquad p_n \geq 0, \qquad \sum_{n} p_n =1.
$$
and are connected with a probability distribution.

The basis $\{ \phi _n \}$ in the space ${\it H}_2$  corresponds to
some observable $\hat G_2$ of the subsystem $S_2$ with discrete
non-degenerate spectrum.  It is determined by the kernel
$$
G_2 (z,z')=\sum_{n} g_n {\phi }_n {\phi }*_n,
\quad g_n = {g*}_n ; \quad g_n \not= g_{n1} \quad if \quad n \not= n1.
$$
The average value of $G_2$ in the state $\rho _2$ is equal to
$$
\int dy \rho _2 (z,z')dz'G(z',z) =
$$
$$
= \sum_{n} g_n \int dy \rho _2 (z,z') dz' \phi _n (z') {\phi _n}^* (z') =
\sum_{n} p_n g_n.
$$
Thus number $p_n$ defines the chance that the observable $\hat
G_2$ has the value $g_n$ in the state $\rho _2$. Obviously, the
kernel $\rho _{n}^{c} (y,y')$ in this case defines the state of
system $S_1$ under condition that the value of variable $G_2$ is
equal to $g_n$. Hence it is natural to call operator $\hat \rho
_{n}^{c}$ as Conditional Density Matrix (CDM) \cite{Kh},
\cite{Sol}
$$
 \hat \rho _{c1|2} = {Tr_2 (\hat P_2 \hat \rho) \over Tr(\hat P_2 \hat
\rho ) }.
$$
It is ({\it conditional}) density matrix for the subsystem $S_1$
under the condition that the  subsystem $S_2$ is selected in a
pure state $\hat \rho _2 =\hat P_2 $.
 It is the most important case for quantum communication. Conditional
density matrix satisfies  all the properties of density matrix.

Conditional density matrix helps to clarify a sense of operations in some
finest experiments.

\section{Examples: System and Subsystems}
\subsection{ Parapositronium}
\noindent
As an example we consider parapositronium, i.e. the system consisting
of an electron and a positron. The total spin of the system is
equal to zero. In this case the nonrelativistic approximation is
valid and the state vector of the system is represented in the
form of the product
$$  {\Psi}({\vec r}_{e},{\sigma}_{e}; {\vec r}_{p}, {\sigma}_{p})
      \quad = \quad
     {\Phi}({\vec r}_{e},{\vec r}_{p})
     \chi({\sigma}_{e},{\sigma}_{p}).
$$
The spin wave function is equal to
$$  \chi({\sigma}_{e},{\sigma}_{p})
      \quad = \quad {1 \over \sqrt{2}}
   ({\chi}_{\vec n}({\sigma}_{e}){\chi}_{-{\vec n}}({\sigma}_{p})
         \quad  - \quad
  {\chi}_{\vec n}({\sigma}_{p}){\chi}_{-{\vec n}}({\sigma}_{e})).
$$
Here ${\chi}_{\vec n}(\sigma)$ and ${\chi}_{(-{\vec n})}(\sigma)$
are the eigenvectors of the operator that projects spin onto the
vector $\vec n$:
$$  ({\vec {\sigma}}{ \vec n})\ {\chi}_{\pm \vec n}(\sigma)
        \quad = \quad \pm
       {\chi}_{\pm \vec n}(\sigma),
$$
The spin density matrix of the system is determined by the
operator with the kernel
$$  \rho({\sigma};{\sigma}^{'})
     \quad = \quad
    {\chi}({\sigma}_{e},{\sigma}_{p})\
    {{\chi}}^{*}({\sigma}^{'}_{e},{\sigma}^{'}_{p}),
$$
The spin density matrix of the electron is
$$  {\rho}_{e}({\sigma},{\sigma}^{'})
     \quad = \quad
  \sum_{\xi}\ {\chi}({\sigma},\xi)\
    {{\chi}}^{*}({\sigma}^{'}, \xi)
     \quad = \quad
$$
$$       {1 \over 2}\
  ({\chi}_{\vec n}(\sigma)\ {\chi}_{(-{\vec n})}({\sigma}^{'})
         \  + \
  {\chi}_{\vec n}(\sigma)\ {\chi}_{(-{\vec n})}({\sigma}^{'}))
      \quad = \quad {1 \over 2}\delta (\sigma - {\sigma}^{'}).
$$
In this state the electron is completely unpolarized.

If an electron passes through polarization filter then
 the pass probability is independent of
the filter orientation. The same fact is valid for the positron if
its spin state is measured independently of the electron.

Now let us consider quite a different experiment. Namely,  the
positron passes through the polarization filter and  the electron
polarization is simultaneously measured. The operator that
projects the positron spin onto the vector $\vec m$ (determined by
the filter) is given by the kernel
$$  P(\sigma,{\sigma}^{'})  \quad = \quad
  {\chi}_{\vec m}(\sigma)\ {\chi}^{*}_{{\vec m}}({\sigma}^{'}).
$$
Now the conditional density matrix of the electron is equal to
$$   {\rho}_{e/p}(\sigma,{\sigma}^{'})
        \ = \
  {\sum_{(\sigma,{\sigma}^{'})}
   {\chi}_{\vec m}(\sigma)\ {\chi}^{*}_{\vec m}({\sigma}^{'})\
       {\chi}({\sigma}_{e},{\sigma}^{'})\
      {{\chi}^{*}}({\sigma}^{'}_{e},\sigma)
      \over
  \sum_{(\xi,\sigma,{\sigma}^{'})}
   {\chi}_{\vec m}(\sigma)\ {\chi}^{*}_{\vec m}({\sigma}^{'})\
       {\chi}(\xi,{\sigma}^{'})\
      {{\chi}^{*}}(\xi,\sigma)}.
$$
The result of the summation is
$$   {\rho}_{e/p}(\sigma,{\sigma}^{'})
         \quad = \quad
   {\chi}_{(-\vec m )}(\sigma)\ {\chi}^{*}_{(-\vec m )}({\sigma}^{'}).
$$

Thus, if the polarization of the positron is selected with
the help of polarizer in the state with well defined spin,
then the electron appears to be polarized in the opposite direction.
Of course, this result is in an agreement with the fact that total
 spin of composite system is equal to zero. Nevertheless
this natural result can be obtained if positron and electron spins
are measured simultaneously.  In the opposite case,  the more
simple experiment shows that the direction of electron and
positron spins are absolutely indefinite.

A.Eistein said "{\it raffinert ist der
Herr Gott, aber boschaft ist Er nicht}".

\subsection{ Quantum Photon Teleportation}
\noindent
In the Innsbruck experiment \cite{Tel} on a photon state teleportation, the
initial state of the system is the result of the unification of
the pair of photons 1 and 2 being in the antisymmetric state
${\chi}({\sigma}_{1},{\sigma}_{2})$ with summary angular momentum
equal to zero and the photon 3 being in the state ${\chi}_{\vec
m}({\sigma}_{3})$ (that is, being polarized along the vector $\vec
m $). The joint system state is given by the density matrix
$$   \rho(\sigma, {\sigma}^{'})
       \quad = \quad
       {\Psi}(\sigma){{\Psi}^{*}}({\sigma}^{'}),
$$
where the wave function of the joint system is the product
$$   {\Psi}(\sigma)
     \quad = \quad
     {\chi}({\sigma}_{1},{\sigma}_{2})\ {\chi}_{\vec m}({\sigma}_{3}).
$$
Considering then the photon 2 only (without fixing the states of
the photons 1 and 3) we find the photon 2 to be completely
unpolarized with the density matrix
$$  {\rho}({\sigma}_{2},{\sigma}_{2}^{'})
       \  = \
    Tr_{(1,3)}\
{\rho}({\sigma}_{1},{\sigma}_{2},{\sigma}_{3};
{\sigma}_{1},{\sigma}_{2}^{'},{\sigma}_{3})
    \  = \
   {1 \over 2}\ \delta ({\sigma}_{2} - {\sigma}_{2}^{'}).
$$
However, if the photon 2 is registered when the state of the
photons 1 and 3 has been determined to be
${\chi}({\sigma}_{1},{\sigma}_{3})$ then the state of the photon 2
is given by the conditional density matrix
$$ {\rho}_{2/\lbrace 1,3 \rbrace}
      \quad = \quad
   {Tr_{(1,3)}\ (P_{1,3}\ {\rho}_{1,2,3}) \over
      Tr\ (P_{1,3}\ {\rho}_{1,2,3})}.
$$
Here $ P_{1,3}$ is the projection operator
$$   P_{1,3} \quad = \quad
      {\chi}({\sigma}_{1},{\sigma}_{3})\ {\chi}^{*}({\sigma}_{1},{\sigma}_{3}).
$$
To evaluate the conditional density matrix it is convenient to
preliminary find the vectors
$$   {\phi}({\sigma}_{1})  \quad = \quad
       \sum_{3}\ {\chi}^{*}_{\vec m}({\sigma}_{3})\ {\chi}({\sigma}_{1},{\sigma}_{3})
$$
and
$$  {\theta}({\sigma}_{2}) \quad = \quad
       \sum_{1}\
       {{\phi}^{*}}({\sigma}_{1})\ {\chi}({\sigma}_{1},{\sigma}_{2}).
$$
The vector $\theta$ equals to
$$  {\theta}({\sigma}_{2}) \quad = \quad
      - {1 \over 2}\ {\chi}_{\vec m}({\sigma}_{2})
$$
and the conditional density matrix of the photon 2 appears to be
equal to
$$ {\rho}_{2/\lbrace 1,3 \rbrace}
      \quad = \quad
      {\chi}_{\vec m}({\sigma}_{2})\ {{\chi}}^{*}_{\vec m}({\sigma}_{2}^{'}).
$$
Thus, if the subsystem consisting of the photons 1 and 3 is forced
to be in the antisymmetric state
${\chi}({\sigma}_{1},{\sigma}_{3})$ (with total angular momentum
equal to zero) then the photon 2 appears to be polarized along the
vector ${\vec m}$.

\subsection{ Entanglement Swapping}
\noindent

In the recent experiment  \cite{Swa} in installation  two pairs of
correlated photons are emerged simultaneously. The state of the
system is described by the wave function
$$   {\Psi}(\sigma) \quad = \quad
   {\Psi}({\sigma}_{1}, {\sigma}_{2}, {\sigma}_{3}, {\sigma}_{4})
            \quad =
\quad {\chi}({\sigma}_{1},{\sigma}_{2}){\chi}({\sigma}_{3},{\sigma}_{4}).
$$
The photons  2 and 3  are selected into  antisymmetric state
$ {\chi}({\sigma}_{2},{\sigma}_{3})$.

What is the state of pair of photons 1 and 4?

Conditional density matrix  of the pair (1-4) is
$$  {\hat \rho}_{14/23} \quad = \quad
  {Tr_{23}({\hat P}_{23}{\hat \rho}_{1234}) \over
  Tr({\hat P}_{23}{\hat \rho}_{1234})},
$$
where operator that selects pair (2-3) is defined by
$$  P_{23}(\sigma, {\sigma}^{'}) \quad = \quad
      {\chi}({\sigma}_{2},{\sigma}_{3})
   {\chi}^{*}({\sigma}_{2}^{'},{\sigma}_{3}^{'})
$$
and density matrix of four photons system is determined by kernel
$$ {\rho}_{1234}(\sigma, {\sigma}^{'}) \quad = \quad
   {\Psi}({\sigma}_{1}, {\sigma}_{2}, {\sigma}_{3}, {\sigma}_{4})
 {\Psi}^{*}({\sigma}_{1}^{'}, {\sigma}_{2}^{'}, {\sigma}_{3}^{'},
    {\sigma}_{4}^{'}).
$$
Direct calculation shows that the pair of the photons (1 and 4)
has to be in pure state with the wave function
$$
\Phi({\sigma}_{1},{\sigma}_{4}) \quad = \quad
      {\chi}({\sigma}_{1},{\sigma}_{4}).
$$
The experiment confirms this prediction.

 \subsection{ Pairs of Polarized Photons}
\noindent

Now consider  a modification of the Innsbruck experiment. Let
there be two pairs of photons $(1,\ 2)$ and $(3,\ 4)$. Suppose
that each pair is in the pure antisymmetric state $\chi$. The
spin part of the density matrix of the total system is given
by the equation
$$  {\rho}(\sigma,{\sigma}^{'})
       \quad = \quad
     {\Psi}(\sigma)\ {{\Psi}^{*}}({\sigma}^{'}),
$$
where
$$  {\Psi}(\sigma)
    \quad = \quad
     {\chi}({\sigma}_{1},{\sigma}_{2})\ {\chi}({\sigma}_{3},{\sigma}_{4}).
$$
If the photons 2 and 4 pass though polarizes, they are polarized along
 ${\chi}_{\vec m}({\sigma}_{2})$ and ${\chi}_{\vec
s}({\sigma}_{4})$
  then the wave function of the system is
transformed into
$$  {\Phi}(\sigma)
      \quad = \quad
 {\chi}_{\vec n}({\sigma}_{1})\ {\chi}_{\vec m}({\sigma}_{2})
\   {\chi}_{\vec r}({\sigma}_{3})\ {\chi}_{\vec s}({\sigma}_{4}).
$$
Here ${\vec n},\  {\vec m}$ and ${\vec r},\  {\vec s}$ are pairs
of mutually orthogonal vectors.

Now the conditional density matrix of the pair of photons 1 and 3
is
$$  {\rho}_{(1,3)/(2,4)}(\sigma,{\sigma}^{'})
     \quad = \quad
        {\Theta}({\sigma}_{1},{\sigma}_{3})\
        {{\Theta}^{*}}({\sigma}_{1}^{'},{\sigma}_{3}^{'}).
$$
The wave function of the pair is the product of wave
functions of each photon with definite polarization
$$   {\Theta}({\sigma}_{1},{\sigma}_{3})
        \quad = \quad
   {\chi}_{\vec n}({\sigma}_{1})\ {\chi}_{\vec r}({\sigma}_{3}) .
$$

We note that initial  correlation  properties of the system appear
only when the photons pass though polarizers. Although the wave
function of the system seems to be a wave function of independent
particles the initial correlation exhibits in correlations of
polarizations for each pair. Pairs of polarized photons appear to
be very useful in quantum communication.

\subsection{ Quantum Realization  of Verman Communication Scheme}
\noindent

Let us recall the main idea of Vernam communication scheme \cite{Ver}.
 In this scheme, Alice encrypts her message (a
string of bits denoted by the binary number $m_{1}$)
 using a randomly generated key $k$.
She simply adds each bit of the message with the corresponding bit
of the key to obtain the scrambled text ($ s = m_{1} \oplus k $,
where $\oplus$ denotes the binary addition modulo 2 without
carry). It is then sent to Bob, who decrypts the message by
subtracting the key  ($s \ominus k = m_{1} \oplus k \ominus k =
m_{1}$). Because the bits of the scrambled text are as random as
those of the key, they do not contain any information. This
cryptosystem is thus provable secure in sense of information
theory. Actually, today this is the only probably secure
cryptosystem!

Although perfectly secure, the problem with this security is that
it is essential that Alice and Bob possess a common secret key,
which must be at least as long as the message itself. They can
only use the key for a single encryption. If they used the key
more than once, Eve could record all of the scrambled messages and
start to build up a picture of the plain texts and thus also of
the key. (If Eve recorded two different messages encrypted with
the same key, she could add the scrambled text to obtain the sum
of the plain texts: $s_{1} \oplus s_{2} = m_{1} \oplus k \oplus
m_{2} \oplus k = m_{1} \oplus m_{2} \oplus k \oplus k = m_{1}
\oplus m_{2}$, where we used the fact that $\oplus$ is
commutative.) Furthermore, the key has to be transmitted by some
trusted means, such as a courier, or through a personal meeting
between Alice and Bob. This procedure may be complex and
expensive, and even may lead to a loophole in the system.

With the help of pairs of polarized photons we can overcome the
shortcomings of the classical realization of Vernam scheme.
Suppose Alice sends to Bob pairs of polarized photons obtained
according to the rules described in the previous section. Note
that the concrete photons' polarizations are set up in Alice's
laboratory and Eve does not know them. If the polarization of the
photon 1 is set up by a random binary number $p_{i}$ and the
polarization of the photon 3 is set up by a number $m_{i} \oplus
p_{i}$ then each photon (when considered separately) does not
carry any information. However, Bob after obtaining these photons
can add corresponding binary numbers and get the number $m_{i}$
containing the information ($m_{i} \oplus p_{i} \oplus
p_{i}=m_{i}$).

In this scheme, a secret code is created during the process of
sending and is transferred to Bob together with the information.
It makes the usage of the scheme completely secure.

\section{Conclusion}
\noindent
 Provided that the  subsystem $S_2$ of composite quantum system $S=S_1 + S_2$
is selected (or will be selected) in a pure state $\hat P_n$
the quantum state of subsystem
$S_1$ is conditional density matrix  $\hat {\rho}_{1c/2n}$. Reduced
density matrix $\hat {\rho }_1$ is connected with conditional
density matrices by an  expansion:
$$
\hat \rho _1  = \sum p_n {\hat \rho}_{1n/2n};
$$
here
$$
 \sum \hat P_n = \hat E, \qquad \sum p_n = 1.
 $$
The coefficients $p_n$ are  probabilities to find subsystem $S_2$
in  pure states $\hat P_n$.

\end{document}